\documentclass[aps,pra,twocolumn,groupedaddress,showpacs,showkeys,superscriptaddress]{revtex4-1}

\usepackage{amsmath}
\usepackage{amssymb}
\usepackage{graphicx}
\usepackage{epstopdf}

\usepackage{subcaption}
\usepackage{mwe}

\usepackage{media9}
\usepackage{hyperref}

\begin{document}

\title{Edge-plasmon assisted electro-optical modulator}

\author{Ivan A. Pshenichnyuk}
\email[correspondence address: ]{i.pshenichnyuk@skoltech.ru}
\affiliation{Skolkovo Institute of Science and Technology, Moscow 121205, Russian Federation}

\author{Gleb I. Nazarikov}
\affiliation{Skolkovo Institute of Science and Technology, Moscow 121205, Russian Federation}

\author{Sergey S. Kosolobov}
\affiliation{Skolkovo Institute of Science and Technology, Moscow 121205, Russian Federation}

\author{Andrei I. Maimistov}
\affiliation{National Research Nuclear University MEPhI (Moscow Engineering Physics Institute), Moscow 115409, Russian Federation}

\author{Vladimir P. Drachev}
\affiliation{Skolkovo Institute of Science and Technology, Moscow 121205, Russian Federation}
\affiliation{University of North Texas, Denton, Texas 76203, USA}

\date{\today}

\begin{abstract}
An efficient electro-optical modulation has been demonstrated here by using an edge plasmon mode specific for the hybrid plasmonic waveguide. Our approach addresses a major obstacle of the integrated microwave photonics caused by the polarization constraints of both active and passive components. In addition to sub-wavelength confinement, typical for surface plasmon polaritons, the edge plasmon modes enable exact matching of the polarization requirements for silicon based input/output grating couplers, waveguides and electro-optical modulators. A concept of the hybrid waveguide, implemented in a sandwich-like structure, implies a coupling of propagating plasmon modes with a waveguide mode. The vertically arranged sandwich includes a thin layer of epsilon-near-zero material (indium tin oxide) providing an efficient modulation at small length scales. Employed edge plasmons possess a mixed polarization state and can be excited with horizontally polarized waveguide modes. It allows the resulting modulator to work directly with efficient grating couplers and avoid using bulky and lossy polarization converters. A 3D optical model based on Maxwell equations combined with drift-diffusion semiconductor equations is developed. Numerically heavy computations involving the optimization of materials and geometry have been performed. Effective modes, stationary state field distribution, an extinction coefficient, optical losses and charge transport properties are computed and analyzed. In addition to the polarization matching, the advantages of the proposed model include the compact planar geometry of the silicon waveguide, reduced active electric resistance $R$ and a relatively simple design, attractive for experimental realization. \end{abstract}

\maketitle

\section{Introduction} \label{sec_intro}

New problems of great practical significance arise as electronic devices like transistors get downscaled to atomic dimensions \cite{datta}. As we seek the ability to engineer materials and devices on an atomic scale, a prediction for the structural, electrical and mechanical properties of new materials, and the rates of chemical reactions become crucial for the transition from nano to molecular electronics \cite{pshenichnyuk-2011, pshenichnyuk-2013}. Attempts to use light to interconnect electronic units at small scales face the mismatch between typical sizes in electronics (tens of nanometers) and communication wavelengths (thousands of nanometers) \cite{liu-2015}. A weak light-matter interaction makes the efficiency of classic opto-electronic devices poor \cite{reed-2010}. Hybrid quasiparticles appear as an elegant approach, which can be used to overcome the existing issues and bring the electronics to the next level. For example, exotic fundamental properties of exciton-polaritons allow to exceed limitations of classical photonics \cite{gibbs-2011}. As it was shown, the exciton-polaritons may form condensates at room temperatures and demonstrate nonlinear coherent behavior typical for quantum fluids \cite{pshenichnyuk-2015, pshenichnyuk-2016, pshenichnyuk-2017, pshenichnyuk-2018}. Some applications of their properties in polariton simulators and potential exciton-polariton integrated circuits have been  discussed recently \cite{berloff-2017, liew-2010, flayac-2013}. 

The technology of subwavelength optics based on surface plasmon polaritons (SPP) is used in numerous applications \cite{gramotnev-2010, barnes-2003}. It is known that SPP modes may combine strong optical confinement and efficient light-matter interaction \cite{liu-2015}. Both properties might improve the performance of active devices for integrated electro-optical systems. Here we suggest a model of efficient and compact electro-optical modulator based on a hybrid plasmonic waveguide (HPWG) concept employing an epsilon-near-zero (ENZ) effect.

While ordinary plasmonic waveguides allow to reach a high degree of optical confinement, they also introduce unavoidable losses, caused by the presence of metallic components. On the other hand, losses in ordinary dielectric waveguides can be made negligible, but the optical mode confinement is obviously much worse. An idea of HPWG is to combine two types of waveguides into a single structure, where a hybrid mode can be formed to compromise losses and confinement \cite{alam-2014, alam-2013}. The strength of the coupling can be controlled by geometrical parameters, such as the distance between the metal surface and the dielectric waveguide.

The ENZ effect, useful for optical switching, appears at infrared frequencies in a certain classes of materials, including transparent conductive oxides (TCO) \cite{naik-2013}. Applying an external electric field an increased concentration of electrons can be achieved, for instance, at a boundary with a dielectric \cite{feigenbaum-2010}. When the critical concentration is reached, the real part of the dielectric permittivity tends to zero in a good correspondence with the Drude theory \cite{caspani-2016}. The ENZ effect was recently demonstrated experimentally in indium-tin-oxide (ITO) \cite{alam-2016}. When the voltage is applied and the real part of permittivity turns into zero, field intensity inside ITO accumulation layer grows in a resonant manner. The imaginary part of the dielectric constant also grows significantly in the ENZ regime, causing a strong dissipation of localized electromagnetic energy in the ITO accumulation layer.

The described concepts in different manifestations were used in many theoretical and experimental research works.
The concept of plasmostor is introduced in the experiments of \citet{dionne-2009}. A silicon based plasmonic waveguide with a thin dielectric layer and silver claddings is used. The applied voltage influences available modes, allowing them to interact constructively or destructively at the output and providing the modulation effect. Doped silicon acts as an active material.
Another model based on a plasmonic slot waveguide is presented in the work by \citet{melikyan-2011}. A dielectric core with a thin layer of ITO is used. The conducting oxide accumulates electrons under the influence of applied voltage, which changes its permittivity and allows to absorb a propagating plasmonic mode. Charge induced changes of the refractive index are more pronounced in ITO than in Si.
A numerical model of ultrasmall fully plasmonic absorption modulator is proposed in \citet{krasavin-2012}.
It is suggested to use a thin plasmonic nanowire with rectangular cross-section separated from a metallic surface by a $5$ nm thick sandwiched ITO/insulator spacer. A compact geometry ($25{\times}30{\times}100$ nm$^3$) and the implementation of high quality dielectric HfO$_2$ allows to make the switching voltage as low as $1$ V. Different gate metals are compared and a better performance of gold is reported.
Further development of the plasmonic slot waveguide modulator appears in experiments of \citet{lee-2014}. In their model ITO fills the space between golden claddings with an additional thin layer of insulator. In such a geometry ENZ regime can be reached at relatively low voltages (${\sim}2$ V), making the model attractive for applications.
Having in mind applications in photonic circuits, where the optical signal is confined in a low loss waveguide, the listed models require conversion of the waveguide mode to the plasmonic slot mode and back, which is accompanied by losses. The classic example of a fully waveguide based absorption modulator is presented in the work of \citet{sorger-2012}. Modulating sandwich, which consists of a thin layers of ITO and an insulator, covered by a golden electrode, is placed on top of a waveguide, forming HPWG. 
More works developing this idea are available \cite{koch-2016,lin-2015}.

There is a family of modulators where ENZ effect is exploited without the implementation of hybrid waveguides. In calculations of \citet{lu-2012} it is suggested to incorporate thin layers of an active material (aluminum zinc oxide) and insulator into the silicon waveguide, without metallic layers and corresponding SPP modes.
Similar approach is suggested in theoretical works \citet{sinatkas-2017} and \citet{qiu-2018}. ITO based capacitor either surrounds the waveguide or partially penetrates inside it. High quality dielectric HfO$_2$ is used in these works, which allows to reach larger charge concentrations at lower voltages. In general ENZ modulators with no plasmonic waveguides implemented are noticeably larger. On the other hand, fully plasmonic solutions with many metallic elements \cite{ayata-2017} introduce significant losses. HPWG with accurately chosen geometry may provide a useful compromise between two approaches. Advanced ways to couple waveguide modes with plasmonic modes may also help to fight plasmonic losses and bring HPWG idea to the next level \cite{haffner-2018}. More details about plasmonic modulators are available in corresponding reviews \cite{amin-2018, liu-2015, dionne-2010}

In current work we remove one of the main obstacles of integrated photonic devices based on HPWG - the polarization constraint. Indeed, plasmons can be excited only by light with the polarization perpendicular to the surface of the metal. Thus, a planar device can only operate with the vertically polarized modes \cite{sorger-2012}. It makes the design incompatible with modern grating couplers, which can provide almost $100$ percent efficiency \cite{michaels-2018}. To match the polarization requirements for grating couplers and the modulator one should develop and use  polarization converters \cite{zhang-2010, majumder-2017, chen-2011, fukuda-2008, pshenichnyuk-2018c}. Obviously, a significant part of the light intensity is lost after the double conversion, before and after the modulation. Besides, it results in additional complexity at the manufacturing side. In this paper we propose a model of a vertically assembled plasmonic modulator employing edge plasmons. Our device works with the same polarization as the couplers and does not require the usage of converters. In contrast with SPP, edge modes have a mixed polarization state, which allows them to serve as coupling providers between a horizontally polarized optical waveguide mode and a metallic electrode. Fundamental properties and dispersion relations of edge plasmons at different boundaries are studied in order to deduce an efficient devise structure. Our modulator has a simple geometry designed for the subsequent experimental realization. The proposed planar structure possess many practical benefits discussed in the text. Technically, to achieve the goal, we propose an advanced HPWG that mixes two edge plasmons with a waveguide mode. The corresponding model is essentially 3D and requires numerically demanding computations. The designed structure provides a comfortable waveguide-plasmon-waveguide conversion length (tunable via geometrical parameters) and allows to reach a required modulation depth.

The paper is organized as following. In Sec.~\ref{sec_theory} we give a short review of a theoretical framework used to perform numerical computations. In Sec.~\ref{sec_edgeplasmons} edge plasmon properties are investigated. The dispersion relation is computed numerically and compared with the SPP dispersion. A polarization state of edge plasmons and their stability with respect to the edge roughness are discussed. Sec.~\ref{sec_design} presents the details of the modulator geometry. 3D computations of the electric field distribution inside the device are presented and discussed. Optical losses are evaluated. High frequency charge transport properties of the modulator are discussed in Sec.~\ref{sec_transport}. The response of the electrons distribution in the active material on the applied voltage is calculated. Optimal parameters for the modulator electrical capacity and related bandwidth are discussed. ENZ wavelength in ITO, as well as the detailed properties of the accumulation layer are investigated. In the last section (Sec.~\ref{sec_offstate}) the on- to off-state transition of the modulator is presented and the extinction coefficient is evaluated. The influence of the modulating sandwich length on the performance of the device is discussed.

\section{Theory}  \label{sec_theory}

Optical model of the modulator is based on Maxwell equations in frequency domain \cite{inan}
\begin{equation}
\nabla \times \mathbf{E}(\mathbf{r}) = +i\omega \mu \mathbf{H}(\mathbf{r}),
\end{equation}
\begin{equation}
\nabla \times \mathbf{H}(\mathbf{r}) = -i\omega \varepsilon \mathbf{E}(\mathbf{r}),
\end{equation}
where $\varepsilon=\varepsilon_r\varepsilon_0$ is the total permittivity (defined as a product of relative and vacuum permittivities) and $\mu$ is a permeability of the medium. The time dependence of electric and magnetic fields is assumed to be  harmonic, i.e. $\mathbf{E}(\mathbf{r},t) = \mathbf{E}(\mathbf{r}) \exp(-i\omega{t})$ and $\mathbf{H}(\mathbf{r},t) = \mathbf{H}(\mathbf{r}) \exp(-i\omega{t})$. For the convenience of the numerical treatment two equations are combined into a single second-order vector wave equation with respect to the electric field
\begin{equation}
\nabla \times \nabla \times \mathbf{E}(\mathbf{r}) - k_0^2 n^2 \mathbf{E}(\mathbf{r}) = 0.
\label{wave_equation}
\end{equation}
Here we introduce a vacuum wave vector $k_0=\omega/c$, a speed of light $c=1/\sqrt{\varepsilon_0\mu_0}$ and a refractive index $n=\sqrt{\varepsilon_r}$ ($\mu=1$ is assumed in this work).
Eq.~\ref{wave_equation} is solved numerically in 3D for a modulator geometry specified in Sec.~\ref{sec_design}. This task is numerically expensive and requires HPC (high performance computing) cluster to be performed. To build the numerical model complex refractive indices (or permittivities) should be fixed for all the materials used in the simulation for a specific wavelength ($\lambda = 1550$ nm is assumed in the paper). Optical parameters for metallic and semiconducting materials are presented in the Tab.~\ref{me_param}.

\begin{table}
  \caption{Optical model parameters \label{me_param}}
  \begin{ruledtabular}
  \begin{tabular}{cccc} 
        & Au & ITO & n-Si  \\ 
    \colrule        
         $n_c$, cm$^{-3}$ & $5\cdot10^{22}$ & $10^{19}$ & $5\cdot10^{17}$  \\         
         $\omega_p$, s$^{-1}$ & $1.3\cdot10^{16}$ & $3\cdot10^{14}$ & $8\cdot10^{13}$  \\           
         $\gamma$, s$^{-1}$ & $3.5\cdot10^{13}$ & $1.7\cdot10^{14}$ & $1.4\cdot10^{13}$  \\           
         $\varepsilon_{\infty}$ & $1$ & $3.9$ & $11.7$\\           
         Re $\varepsilon(\omega_0)$ & $-106.7$ & $3.8$ & $11.7$ \\ 
         Im $\varepsilon(\omega_0)$ & $3.1$ & $0.008$ & $5\cdot10^{-5}$  \\                             
         Re $n(\omega_0)$  & $0.2$ & $2.0$ & $3.4$  \\  
         Im $n(\omega_0)$  & $10.3$ & $0.002$ & $7\cdot10^{-6}$ \\  
    \colrule                  
         References & \cite{olmon-2012}  & \cite{kulkarni-1996,sinatkas-2017,michelotti-2009} & \cite{arora-1982,cleary-2010,sinatkas-2017}  \\                                                        
  \end{tabular}
  \end{ruledtabular}
\end{table}

Along with full 3D numerical computations we use the mode analysis technique. Considering a wave propagating in $x$-direction and confined in $y$ and $z$ directions, one may write
\begin{equation}
\mathbf{E}(\mathbf{r}) = \mathbf{E}(y,z)e^{i\beta{x}},
\label{3dto2d}
\end{equation}
where the propagation constant $\beta\equiv n_{\text{eff}}k_0$, expressed through effective mode indices $n_{\text{eff}}$, is introduced.  Such a substitution (Eq.~\ref{3dto2d}) makes Eq.~\ref{wave_equation} effectively 2D and, thus, allows to perform fast calculations and obtain useful results without the necessity to use supercomputers. Effective mode indices $n_{\text{eff}}$ and corresponding field distributions $\mathbf{E}(y,z)$ are computed numerically using ordinary PC.
In our model (see Sec.~\ref{sec_design}) the cross-section of the modulator has a rather complicated structure containing many areas and the permittivity distribution $\varepsilon=n^2(y,z)$ should be considered as a function of $y$ and $z$. Therefore, Eq.~\ref{wave_equation} under condition \ref{3dto2d} in a scalar form prepared for the numerical treatment reads:
\begin{equation}
\frac{\partial^2E_x}{\partial{y}^2} + \frac{\partial^2E_x}{\partial{z}^2} + k_0^2n^2E_x = 
 i\beta \frac{\partial E_y}{\partial{y}} + i\beta \frac{\partial E_z}{\partial{z}},
\end{equation}
\begin{equation}
\frac{\partial^2E_y}{\partial{z}^2} + (k_0^2n^2-\beta^2)E_y = 
i\beta\frac{\partial{E_x}}{\partial{y}} +  \frac{\partial^2E_z}{\partial{y}\partial{z}},
\end{equation}
\begin{equation}
\frac{\partial^2E_z}{\partial{y}^2} + (k_0^2n^2-\beta^2)E_z = 
i\beta\frac{\partial{E_x}}{\partial{z}} + \frac{\partial^2E_y}{\partial{z}\partial{y}} .
\end{equation}
The propagation constant $\beta$ should be considered as a discrete eigenvalue here.

To investigate charge transport in the modulator, semiconductor drift-diffusion system of equations for the electron density $n=n(\mathbf{r},t)$ and potential $\varphi=\varphi(\mathbf{r},t)$ in the form \cite{piprek,yuan}
\begin{equation}
\nabla\cdot(\varepsilon_0\bar{\varepsilon}\nabla\varphi) = e(n-N_d),
\label{eq_dd1}
\end{equation}
\begin{equation}
\frac{{\partial}n}{{\partial}t} = \nabla\cdot (D_n\nabla{n}-n\mu_n\nabla(\varphi+\chi)),
\label{eq_dd2}
\end{equation}
is solved numerically. Here $\bar{\varepsilon}$ is a static permittivity \cite{kittel}, $N_d$ - doping level, $\mu_n$ - mobility of electrons, $\chi$ - electron affinity and $D_n$ - diffusion coefficient. Since both ITO and silicon are n-doped, only donor type conductance is considered. Model parameters for silicon and ITO are collected in Tab.~\ref{dd_param}.

In certain regimes, when the Maxwell-Boltzmann statistics is applicable, diffusion coefficient can be expressed simply as $D_n=\mu_nk_bT/e$. In our computations the complete Fermi-Dirac statistics is taken into account, and the diffusion coefficient reads \cite{sinatkas-2017}
\begin{equation}
D_n = \frac{\mu_nk_bT}{e}\frac{F_{1/2}(\eta)}{F_{-1/2}(\eta)},
\end{equation}
with
\begin{equation}
\eta = F_{1/2}^{-1}(n/N_c),
\end{equation}
where $F_{\pm1/2}$ and $F^{-1}_{1/2}$ are direct and inverse Fermi-Dirac integrals of the order $\pm1/2$. Effective density of states in the conduction band is expressed as \cite{kittel}
\begin{equation}
N_c = 2 \left( \frac{m^{\ast}k_bT}{2\pi\hbar^2}  \right)^{3/2}.
\end{equation}
Reduced masses of electrons $m^{\ast}$ for ITO and n-Si are shown in Tab.~\ref{dd_param}. The room temperature $T=300$ K is assumed in computations.

The Drude theory links the charge density evolution under the applied voltage with the optical model.
Drude based frequency domain permittivity \cite{novotny, maier}
\begin{equation}
\varepsilon(\omega) = \varepsilon_{\infty} - \frac{\omega_p^2}{(\omega^2 + \gamma^2)}
+ i\frac{\gamma\omega_p^2}{\omega(\omega^2 + \gamma^2)},
\label{drude}
\end{equation}
describes well optical properties of ITO and n-Si in C+L telecom range ($1530 - 1625$ nm). Here $\omega_p=\sqrt{n_ce^2/(\varepsilon_0m^{\ast})}$ is the plasmonic frequency and $\gamma = e/(\mu_nm^{\ast})$ is the relaxation coefficient. Asymptotic material permittivity at large frequencies is denoted as $\varepsilon_{\infty}$ and $n_c$ stands for the charge carriers concentration.
Inhomogeneous concentration profile $n(\mathbf{r})$ obtained from Eqs.~\ref{eq_dd1} and \ref{eq_dd2} can be substituted here to obtain the distribution of permittivity. In particular, the structure of the accumulation layer at the boundary between ITO and dielectric can be computed with high precision and used in the optical model, as shown in Sec.~\ref{sec_offstate}. Drude based permittivities and refractive indices, used in optical computations are collected in Tab.~\ref{me_param}. They are computed at the frequency $\omega_0 = 1.215\cdot10^{15}$ s$^{-1}$ (corresponding to the wavelength $\lambda_0=1550$ nm).  Refractive indices of dielectrics, namely HfO$_2$ and SiO$_2$,  are assumed to be constant and real in C+L range with $n=2.0$ and $n=1.97$ respectively \cite{alkuhaili-2004, wilk-2001,robertson-2004}.

\begin{table}
  \caption{Drift-diffusion equations parameters \label{dd_param}}
  \begin{ruledtabular}
  \begin{tabular}{ccc} 
         & ITO & n-Si \\ 
    \colrule
        $m^{\ast}/m_{e}$  & $0.35$ & $0.25$  \\ 
        $\mu_n$, cm$^2$/(V$\cdot$s) & $30$ & $500$  \\ 
         $\chi$, V &  $4.8$ & $4.05$  \\         
         $\bar{\varepsilon}$ &  $8.9$ & $11.7$  \\         
         $N_d$, cm$^{-3}$  & $10^{19}$ & $5.0\cdot10^{17}$  \\         
    \colrule                  
         References & \cite{kulkarni-1996,sinatkas-2017,michelotti-2009} & \cite{arora-1982,cleary-2010,sinatkas-2017} \\                                                        
  \end{tabular}
  \end{ruledtabular}
\end{table}

Numerical experiments are performed using the commercial software Comsol Multiphysics 5.3a with additional modules WaveOptics and Semiconductors. HPC cluster of Skoltech called Pardus is used for calculations.

\section{Edge plasmons} \label{sec_edgeplasmons}

\begin{figure*}
\centerline{\includegraphics[width=0.95\textwidth]{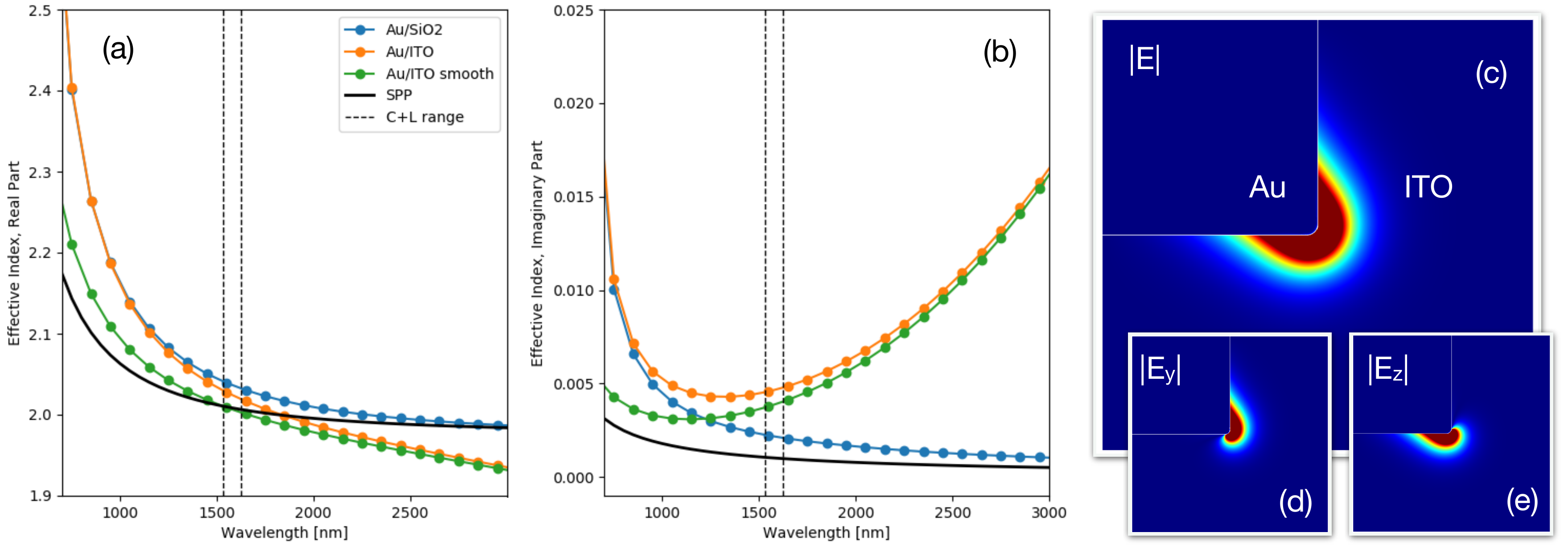}}
\caption {Properties of edge plasmons. Dispersion relations are plotted as a dependence of the effective mode indices on wavelength $n_{eff}(\lambda)$. Both real part (a) and imaginary part (b) are shown. A family of curves is plotted to compare ordinary SPP, edge plasmons at the gold/silica and gold/ITO interfaces. Smooth edge (radius 300 nm) vs sharp edge plasmons comparison is shown. Vertical dashed lines highlight C+L range. The absolute value of the electric field distribution is shown in (c), while two transverse projection $|E_y|$ and $|E_z|$ are shown in (d) and (e) respectively. \label{fig_01}}
\end{figure*}

Before the introduction of the modulator and its parameters (see the next section) we review some important fundamental properties of edge plasmons. As it is mentioned in the introduction, these modes provide a useful alternative to SPP in the case of  plasmonic modulators. In contrast to SPP they have a symmetric mixed polarization state, which allows to interact with both polarizations of the waveguide mode. A field distribution of the edge mode at the ITO/Au boundary is shown in Fig.~\ref{fig_01} (c). The plasmon propagates in the $x$-direction (perpendicular to the surface of the picture). Projections $E_y$ and $E_z$ of the field on the transverse directions, revealing the polarization state, are shown separately (Fig.~\ref{fig_01} (d)-(e)). 

Dispersion relations of edge plasmons are shown in Fig.~\ref{fig_01} (a) and (b). Instead of usual dependence of angular frequency on a wave vector $\omega(k)$, real and imaginary parts of $n_{\text{eff}}(\lambda)$ are plotted (as defined in the previous section). To go back to the usual notations one may use $k=n_{\text{eff}}\,\omega/c$ and $\omega=2{\pi}c/\lambda$, where $c$ is a speed of light and $\lambda$ is a wavelength. While the real part of $n_{\text{eff}}$ is associated with the mode propagation, the imaginary part reflects its decay. A family of curves describing plasmons excited at the golden edge surrounded by SiO$_2$ or ITO (see below) is presented. The solid black line shows the dispersion relation of SPP given by the analytical formula \cite{novotny}
\begin{equation}
k = \frac{\omega}{c}\sqrt{\frac{\varepsilon_1(\omega)\varepsilon_2}{\varepsilon_1(\omega)+\varepsilon_2}},
\label{dispersion_spp}
\end{equation}
where $\varepsilon_2 = 3.9$ is a permittivity of SiO$_2$ and $\varepsilon_1(\omega)$ is associated with gold and expressed by the Drude formula (Eq.~\ref{drude}). C+L telecom range ($1530 - 1625$ nm), crucial for  applications,  is shown in Fig.~\ref{fig_01} (a)-(b) using vertical dashed lines.

In our modulator design plasmons are excited at the Au/ITO boundary.  The advantages of the selected sandwich structure, materials and their order in the sandwich, are discussed in detail in Sec.~\ref{sec_transport}.The difference between the Au/ITO boundary and frequently used Au/SiO$_2$ boundary \cite{sorger-2012} can be understood with a help of the dispersion relations in Fig~\ref{fig_01} (a)-(b) (orange and blue lines respectively). Au/ITO edge plasmons are slightly more lossy than Au/SiO$_2$ plasmons at $\lambda=1550$ nm (Fig.~\ref{fig_01} (b)). That is a consequence of the fact that ITO permittivity has a nonzero imaginary part. However, gently doped ITO films remain mostly transparent at near-infrared and electro-magnetic losses are low \cite{fang-2014,wei-2017}. It is evident from the picture that the edge plasmons in ITO decay fast at larger wavelengths, interacting stronger with the medium. One can see very different asymptotics of the orange line and the blue line. Both curves demonstrate growth at small wavelengths. An interplay of two trends produces a point with minimal losses for Au/ITO plasmons which appears at $1250$ nm. Losses in C+L range, which is slightly shifted with respect to the minimum, are still quite low. Additional tuning can be performed by changing the doping level in ITO with limitations discussed in Sec.~\ref{sec_transport}. Real parts of ITO permittivity and SiO$_2$ permittivity have the values close to each other, which results in a family of similar curves in Fig.~\ref{fig_01}(a). The SPP case is shown for comparison (black solid curve).

In the experimental implementation metallic edges cannot be perfectly sharp. To investigate the stability of edge modes with respect to the edge roughness we perform the comparison between sharp edge solutions and rounded edge solutions (the radius of the rounded corner is 300 nm), which is shown in Fig.\ref{fig_01} (a) and (b) with orange and green curves respectively. For long-wavelength plasmons the difference is negligible, since the plasmon characteristic sizes are much larger than the curvature radius. Solutions are also similar around $\lambda=1550$ nm. At small wavelengths the difference can be significant. When the size of the plasmon becomes smaller than the curvature radius it perceives the edge rather like a surface and its dispersion curve moves closer to SPP (see how the green line approaches the black SPP line at small $\lambda$ in Fig.~\ref{fig_01} (a)). It is interesting to note, that, due to this effect, rounded edge plasmon losses in ITO can be even smaller than losses of edge plasmons in SiO$_2$  (compare blue and green curves in Fig.~\ref{fig_01} (b) below $1000$ nm).

\section{Modulator design and field distribution} \label{sec_design}

\begin{figure}
\centerline{\includegraphics[width=0.45\textwidth]{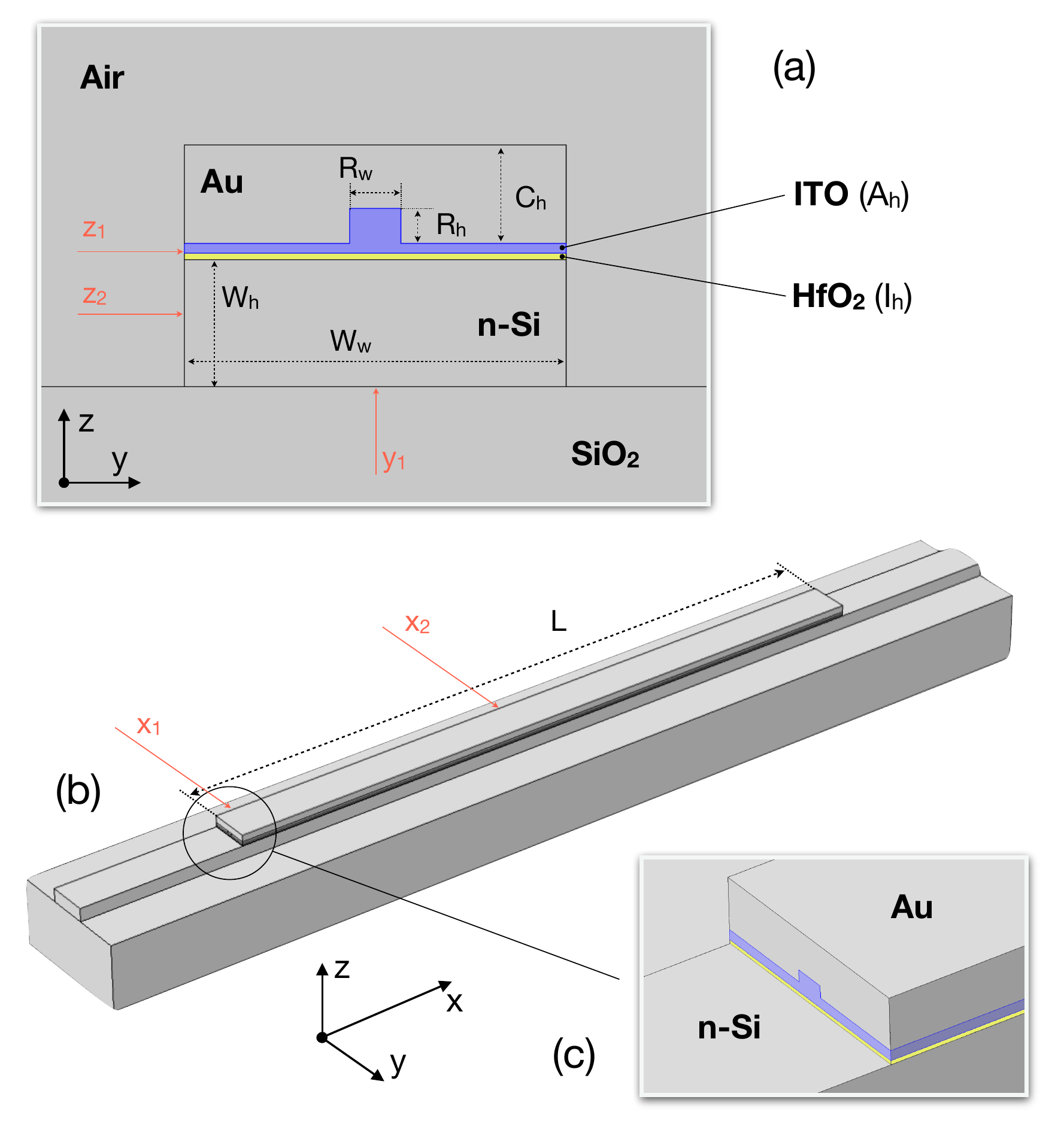}}
\caption {Structure and geometry of the modulator. (a) cross section of the modulator (b) 3D model used in numerical computations (c) the part, where the waveguide enters the modulator. ITO and HfO$_2$ are shown with violet and yellow colors respectively. Cross sections used throughout the paper for the visualization of the field distribution are shown with thin red lines. The following geometrical parameters were implemented in the computations: $W_w=600$ nm, $W_h = 200$ nm, $I_h = 10$ nm, $A_h = 15$ nm, $R_w = 80$ nm, $R_h = 55$ nm, $C_h = 155$ nm, $L = 6845$ nm. \label{fig_02}}
\end{figure}

\begin{figure}
\centerline{\includegraphics[width=0.45\textwidth]{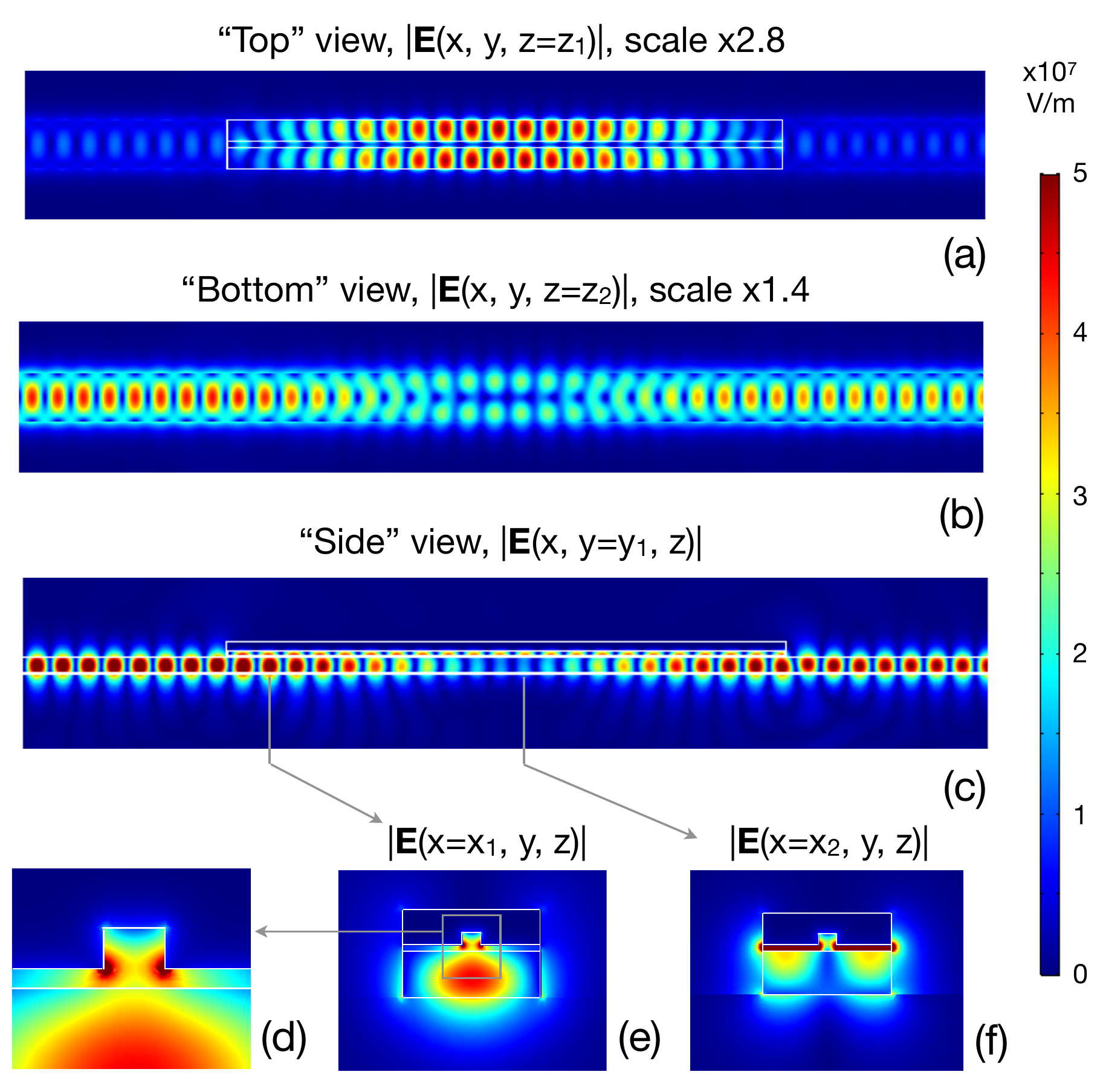}}
\caption {Distribution of the absolute value of the electric field $|\mathbf{E}(x,y,z)|$ inside the modulator when the voltage is not applied and the transmission is maximal ('on-state'). Five different cross-sections of 3D model are shown: (a) horizontal plane which shows the field inside the sandwich, (b) horizontal cross-section which passes through the center of the waveguide, (c) vertical cross-section also shows the field inside the waveguide, (e) transverse cross-section which show the distribution at the entrance of the modulator, (f) the field distribution in the central part of the modulator. (d) the waveguide fundamental mode coupled to two edge plasmons. It is evident how the waveguide mode transforms into the plasmonic mode and back. \label{fig_03}}
\end{figure}

The proposed modulator design is presented in Fig.~\ref{fig_02}. Single mode silicon waveguide (600 nm x 200 nm) is covered with the modulating sandwich consisting of few layers: high quality dielectric HfO$_2$ (10 nm), ITO (15 nm) and golden contact (155 nm). Additional structural element, which we call the plasmonic rail (or just rail), is placed on top of the ITO layer (see Fig.~\ref{fig_02} (a)). The resulting geometry contains two Au edges, which support plasmons, as described in the previous section. Such modes can be coupled to the $y$-polarized waveguide mode. On the contrary, a flat structure without the rail supports only $z$-polarized SPP modes which do not interact with the waveguide mode. It is relatively easy to fabricate such a geometry which makes the resulting modulator design promising for applications. Certain deviations from the specified geometry of the rail are acceptable. The precision of vertical lines, for example, does not have to be very strict, since edge plasmons are quite stable to such variations. The edges can also be smoothed, as it was shown in the previous section. Since the rail brakes the symmetry in $y$-direction, 2D computations would not be sufficient for such a device and full 3D modeling is required.

The size of the plasmonic rail ($80$ nm $\times$ $55$ nm in our model) is chosen to maximize the modulator performance, which means at least several things. HPWG should provide an effective conversion of the waveguide mode to the plasmon (which can be efficiently modulated) and back. In the on-state, when the voltage is not applied, conversion losses should be minimal.
According to our computations, conversion between modes occurs in a manner of beats. Such a behavior is typically expected when two or more interacting modes exist in the system. Even in a simplest case HPWG contains at least two modes, which appear as a result of hybridization of one waveguide mode and one plasmonic mode \cite{sorger-2012}. The analysis for our edge modulator reveals three modes (two edge plasmons hybridized with one waveguide mode), but just two of them are excited. Strong interaction between modes, caused by small distances and large confinement, results in a strong mixing. In contrast with directional couplers (and similar devices), where the coupled mode theory can be used to describe weak interaction between modes, there is no simple analytical theory describing HPWG \cite{alam-2014}. For this reason different geometries of the rail, along with other parameters of the modulator, are tested numerically in order to obtain the configuration with suitable insertion losses and beats period. 
Another factor, which defines the performance, is related to the off-state. When the voltage is applied the signal attenuation has to be maximal. It should be also taken into account that the accumulation layer in ITO, being formed, changes the character of interaction between modes. The final geometry of the modulator is defined by both on- and off-states. The latter one is discussed in Sec.~\ref{sec_offstate}.

The absolute value of the electric field $|\mathbf{E}(x,y,z)|$ inside the modulator is shown in Fig.~\ref{fig_03}. To visualize a 3D distribution, five different cross-sections are plotted. The position and orientation of the selected cross-sections is denoted with thin redlines in Fig.~\ref{fig_02}.  In Fig.~\ref{fig_03} (a) 'horizontal' cross-section $|\mathbf{E}(x,y,z=z_1)|$ is fixed at the center of the gap between the silicon waveguide and the electrode. In Fig.~\ref{fig_03} (b) the cross-section $|\mathbf{E}(x,y,z=z_2)|$ shows the field in the center of the waveguide. It is evident from the pictures that the waveguide mode, passing through the sandwich is smoothly transformed into the plasmon and back. The 'side' view $|\mathbf{E}(x,y=y_1,z)|$ in Fig.~\ref{fig_03} (c) illustrates the same process. Since the modes inside the modulator are mixed by strong interaction, we recognize a plasmon by the high value of the field intensity near the golden surface. Note the scaling factor, which is different in Fig.~\ref{fig_03} (a) and (b), and was introduced for the better visualization. Fig.~\ref{fig_03} (e) and (f) show the transverse cross-section that illustrate the distribution of the field at the entrance of the modulator, and in the center, where the plasmonic state is most populated. One can see that the metal edges act as coupling providers (see Fig.~\ref{fig_03} (d)) for the $y$-polarized waveguide mode and allow to concentrate the field in the plasmonic gap (Fig.~\ref{fig_03} (f)), where it can be efficiently modulated.

Geometrical parameters of the plasmonic rail, in particular its width $R_w$, can be used to tune the interaction between modes, which influences the period of beats. In the proposed model (Fig.~\ref{fig_02}) the period is equal to $6.8 \,\mu$m and the length $L$ of the modulating sandwich is chosen accordingly. If these lengths are not synchronised, losses at the output of the modulator, related to the mode conversion (namely, the conversion of the hybrid mode into the waveguide mode), are generally larger. It is suggested that the sandwich length must be equal to an integer number of beats periods. An example of the devise with a double length is presented in the last section. However, multiple conversions of the signal to the plasmon and back are also connected with additional losses. According to our computations, $6.8 \,\mu$m long sandwich, containing just one period, allows to reach an acceptable modulation depth as it is discussed in Sec.~\ref{sec_offstate}. For the selected geometry the transmission coefficient (the ratio of the output and input intensities $I_{\text{out}}/I_{\text{in}}$) $T=0.747$ corresponds to the optical losses coefficient
\begin{equation}
\kappa_{\text{ol}}=10\log_{10}{\frac{I_{\text{in}}}{I_{\text{out}}}}=1.27 \, \text{dB}.
\label{optical_losses}
\end{equation}
The second important geometrical parameter of the plasmonic rail is its height $R_h$. If it is too large, an additional gap mode appears which interacts with existing modes and makes the picture more complicated. In our geometry ($55$ nm height of the rail) this mode is attenuated.

\section{Switching} \label{sec_transport}

\begin{figure}
\centerline{\includegraphics[width=0.45\textwidth]{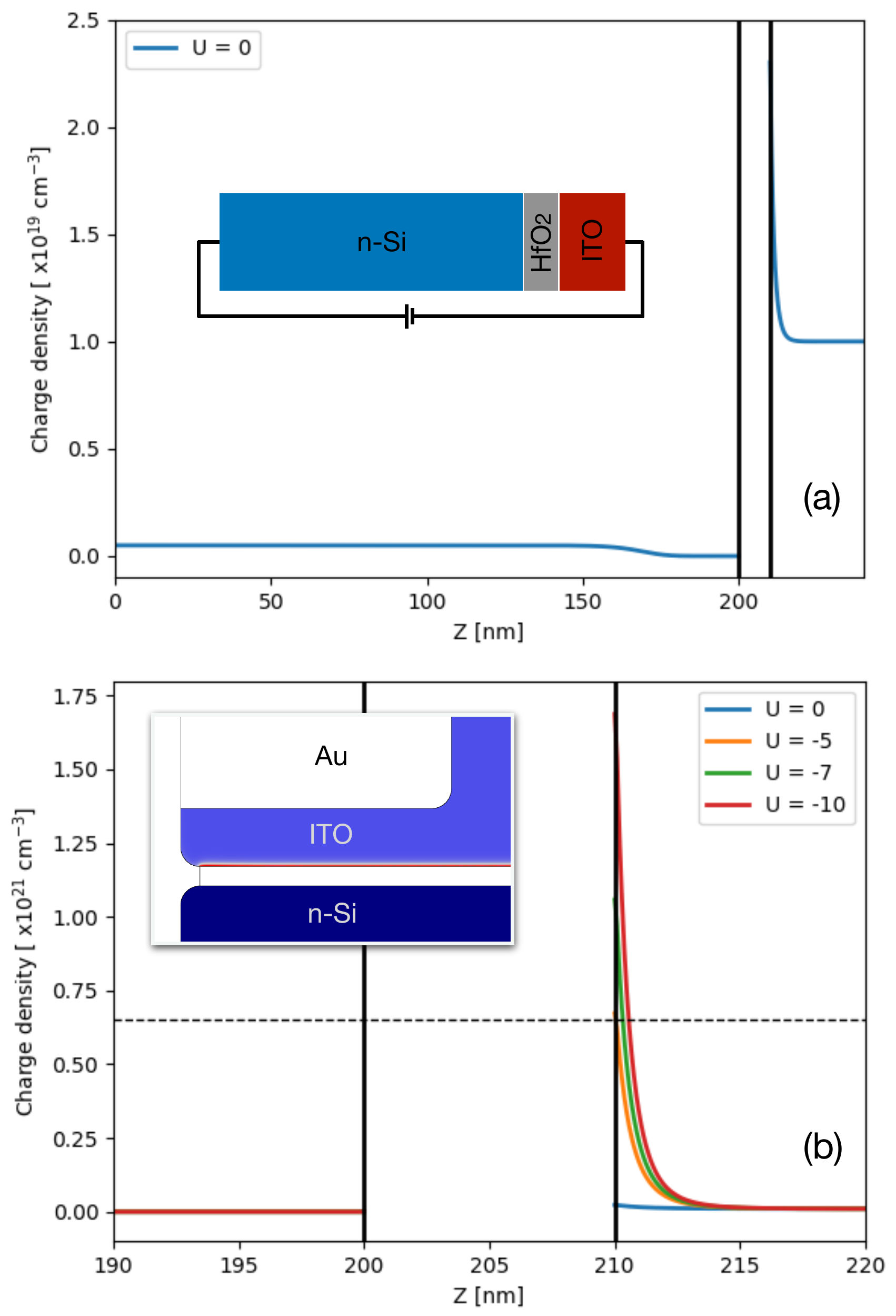}}
\caption {Charge distribution in the modulator (a) when the external voltage is switched off and the concentration profiles are determined by initial doping level and electron affinities of materials (b) when voltage is applied (three different voltages are shown). Mind 100 times scale difference between (a) and (b). Solid vertical lines show the boundaries between materials. Horizontal dashed line marks the value of resonant concentration $n_c^{enz}$. The insets show the structure of the sandwich and 2D charge distribution, where the accumulation layer is shown with red. \label{fig_04}}
\end{figure}

To understand the state switching in the modulator and design the optimal structure of the sandwich, the detailed drift-diffusion analysis based on Eq.~\ref{eq_dd1} and Eq.~\ref{eq_dd2} is performed. Main results are summarized in Fig.~\ref{fig_04}. Charge density evolution in the modulator is symmetric in the $x$-direction. Despite the absence of the symmetry in the $y$-direction, caused by the plasmonic rail, one dimensional charge transport models provide a precise description of the accumulation layer at the ITO/HfO$_2$ boundary. For drift-diffusion computations the rail has been ignored and one dimensional models used to obtain the charge concentration profiles. We also perform 2D computations in some regimes for the comparison, where the rail was modeled explicitly (see the inset in Fig.~\ref{fig_04} (b)).

In order to develop an essential understanding of the modulating sandwich one should think about it as a nano-capacitor. In our model the capacitor is produced by the combination of a doped silicon, HfO$_2$ and ITO (see the inset in Fig.~\ref{fig_04} (a)). The Au contact is attached to ITO. It is not explicitly modelled in the drift-diffusion approach, but enters the equations through the boundary conditions. Silicon is assumed to be grounded. The thin layer of hafnium oxide acts as an insulator. Electron densities in silicon and ITO are $5{\cdot}10^{17}$ cm$^{-3}$ and $10^{19}$ cm$^{-3}$ respectively, which allows them to conduct current. Note that ITO films can be doped up to the level $10^{21}$~$cm^{-3}$ \cite{kulkarni-1996}. If the voltage is applied, the charge is accumulated at the capacitor. When the critical charge concentration is reached, ENZ effect is initiated in ITO leading to the strong attenuation of the optical signal. It is easy to evaluate the critical concentration from Eq.~\ref{drude}, assuming the real part is equal to zero:
\begin{equation}
n_c^{enz} = \frac{m^{\ast}}{e^2}\varepsilon_{\infty}\varepsilon_0 (\omega^2 + \gamma^2) = 6.5\cdot 10 ^{20} \text{cm}^{-3}
\label{enz_concentration}
\end{equation}
($\lambda=1550$ nm is assumed). Having in mind the classic formula for a capacitor $Q=CU$, where $C$ is a capacitance, $U$ - voltage and $Q$ - accumulated charge, one may conclude that since $Q$ is fixed by Eq.~ \ref{enz_concentration}, and we want $U$ to be small (for practical applications), the capacitance, on the contrary, should be large. To increase $C$ one may decrease the thickness of the insulating layer, which is practically difficult and potentially allows electrons to tunnel through the capacitor as, for example, in the floating gate transistor, producing undesired effects. Another way to increase $C$ is to use high a quality dielectric with a large static permittivity, like HfO$_2$ with $\bar{\varepsilon}=25$  \cite{robertson-2004} \cite{wilk-2001}. In our model, $10$ nm thick layer of the hafnium oxide allows to keep the voltage below $10$ V. Based on this analysis one should avoid also structures with multiple layers of insulator, since the capacitance of such junction is decreased. Of course, if $C$ is increased, the charging time also becomes larger and makes the modulation process slower. Thereby, there is a certain trade between the modulation speed and power consumption.

Having in mind a sandwich with a single insulating layer, an active material (ITO) and a metallic contact, two possible configurations can be considered. Instead of Si/HfO$_2$/ITO/Au structure used in our model, alternative Si/ITO/HfO$_2$/Au sandwich can be used. In such a configuration ITO is in a direct contact with silicon, the capacitor is  formed between ITO and Au and plasmons are excited at the boundary between Au and HfO$_2$. Since Si and ITO have different electron affinities, there is a contact potential difference at the Si/ITO boundary and the charge density distribution function has an extra peak there. It produces undesirable refractive index gradients in the modulator. It is also hard to model this peak, since its amplitude depends on the details of the fabrication. On the other hand, the described effect is noticeable only at low voltages and becomes small at high voltages (like the modulator switching voltage, around $8$ V). From this point of view both types of sandwiches can be used. In our model we prefer to use Si/HfO$_2$/ITO/Au structure, minimizing uncertainty in parameters and possible parasitic effects.

The charge density distribution in the modulator (along $z$-axis) at the voltage switched off is plotted in Fig.~\ref{fig_04} (a). Boundaries of materials are shown with vertical black lines. The structure of the sandwich is demonstrated in the inset. It is clear from the picture, that  there is a small charge at the capacitor even at zero voltage. It is a consequence of difference between electron affinities of ITO and Si which results in a small potential difference (around $0.75$ V). Asymptotical values of the charge distribution far from the capacitor correspond to the doping level of materials. To switch the state of the modulator the voltage is applied to the golden contact. Electron density profiles are presented for voltages $-5$ V, $-7$ V and $-10$ V in Fig.~\ref{fig_04} (b) (note the scale difference with Fig.~\ref{fig_04} (a)). The horizontal dashed line shows the ENZ value of the concentration (Eq.~\ref{enz_concentration}) that should be reached for the efficient optical modulation. For a given capacitance $C_0$, the critical value is reached starting from voltage $-5$ V, which is low enough to be practical. Using the data from Fig.~\ref{fig_04} (b) it is easy to evaluate $C_0$ as
\begin{equation}
C_0 = \frac{eS_{mod}}{U} \int\limits_{z{\in}ITO} (n_c^U(z)-n_c^0(z))dz = 0.1 \,\text{[pF]},
\label{cap_eval}
\end{equation}
where $S_{mod}=6\cdot10^6\,\text{nm}^2$ is a transverse area of the modulating sandwich and the integral is taken through the ITO lead of the capacitor, to evaluate the accumulated charge in the active material. Further, $n_c^U$ is the charge density profile at the voltage $U$ and $n_c^0$ is the profile for $U=0$ (blue line in Fig.~\ref{fig_04}). The classic formula for a parallel-plate capacitor applied to our nano-capacitor gives very close values
\begin{equation}
C_0 \approx \varepsilon_0\bar{\varepsilon}\frac{S_{mod}}{d},
\label{cap_eval_classic}
\end{equation}
where $d$ is a thickness of HfO$_2$. It is clear from this formula that by changing the insulating material from HfO$_2$ to SiO$_2$ the capacitance becomes approximately $6$ times smaller and requires $6$ times larger voltage to switch the modulator. The lack of capacitance can be compensated by decreasing the thickness of the insulator from $10$ nm to $1.7$ nm which is not practical for fabrication. Thus, HfO$_2$ remains the main candidate for the insulating material in the proposed device.

One important characteristic number obtained from the drift-diffusion model is the thickness of the accumulation layer in ITO. According to our computations (Fig.~\ref{fig_02} (b)) this layer is quite thin. The specific number can be defined using the equation
\begin{equation}
\frac{1}{t} \int\limits_{z_0}^{z_0+t} n_c^U(z) dz = n_c^{enz},
\label{acc_layer_thickness}
\end{equation}
where $z_0$ is a coordinate of HfO$_2$/ITO boundary and $t$ is a thickness of the active layer. The numerical evaluation procedure returns the value $t\sim1$ nm. Thus, independently on how thick the ITO layer is, the active layer is very thin. The layout of the accumulation layer in 2D computations is also shown in Fig.~\ref{fig_04} (b) (the inset). It corresponds to the thin red line at the ITO/insulator boundary. Charge density in the insulator is equal to zero (white color), and the charge density in the Au contact is not modeled explicitly, but added as a boundary condition in drift-diffusion calculations. The fact that the accumulation layer is very thin is taken into account in optical calculations that require an advanced resolution in this area of the model (see the next section).

Equations \ref{eq_dd1} and \ref{eq_dd2} are solved in both time-dependent and stationary regimes (when $\partial{n}/\partial{t}=0$). The formation of the accumulation layer, depicted in Fig.~\ref{fig_04}, is also studied dynamically. According to the computations, the formation time of the layer is $\Delta{t} \sim 1$~ps, which corresponds to the modulator bandwidth $\Delta{f} \sim 1$ THz. On the other hand, the bandwidth of the modulator is defined by the charging time of the nano capacitor. Classically, this time is of the order of $\Delta{t}\sim RC_0$. To evaluate the active resistance of the leads $R$ one can use the formula $R=l {\sigma}^{-1} S_{mod}^{-1}$, where $\sigma=n_c\mu_ne$ is the conductivity and $l$ is the conductor length (in $z$-direction). The active resistance of our modulator is defined by the resistance of the silicon waveguide, mainly because of its transverse size (the resistance of 'Si lead of the capacitor' is roughly one order of magnitude larger, than the resistance of 'ITO lead'). Using the value of $R$ for doped silicon to evaluate the charging time gives the value $RC_0 \sim 1$~ps which well coincides with our numerical result. Consequently, the doping level of silicon directly influences the modulator bandwidth and increasing the concentration of electrons 10 times one may obtain 10 times wider band. On the other hand the larger doping leads to the increased imaginary part of permittivity in silicon, which attenuates the optical signal, propagating in such a material. Thus, a compromise between the waveguide losses and the modulator bandwidth should be reached.

\section{Modulator off-state}  \label{sec_offstate}

\begin{figure}
\centerline{\includegraphics[width=0.45\textwidth]{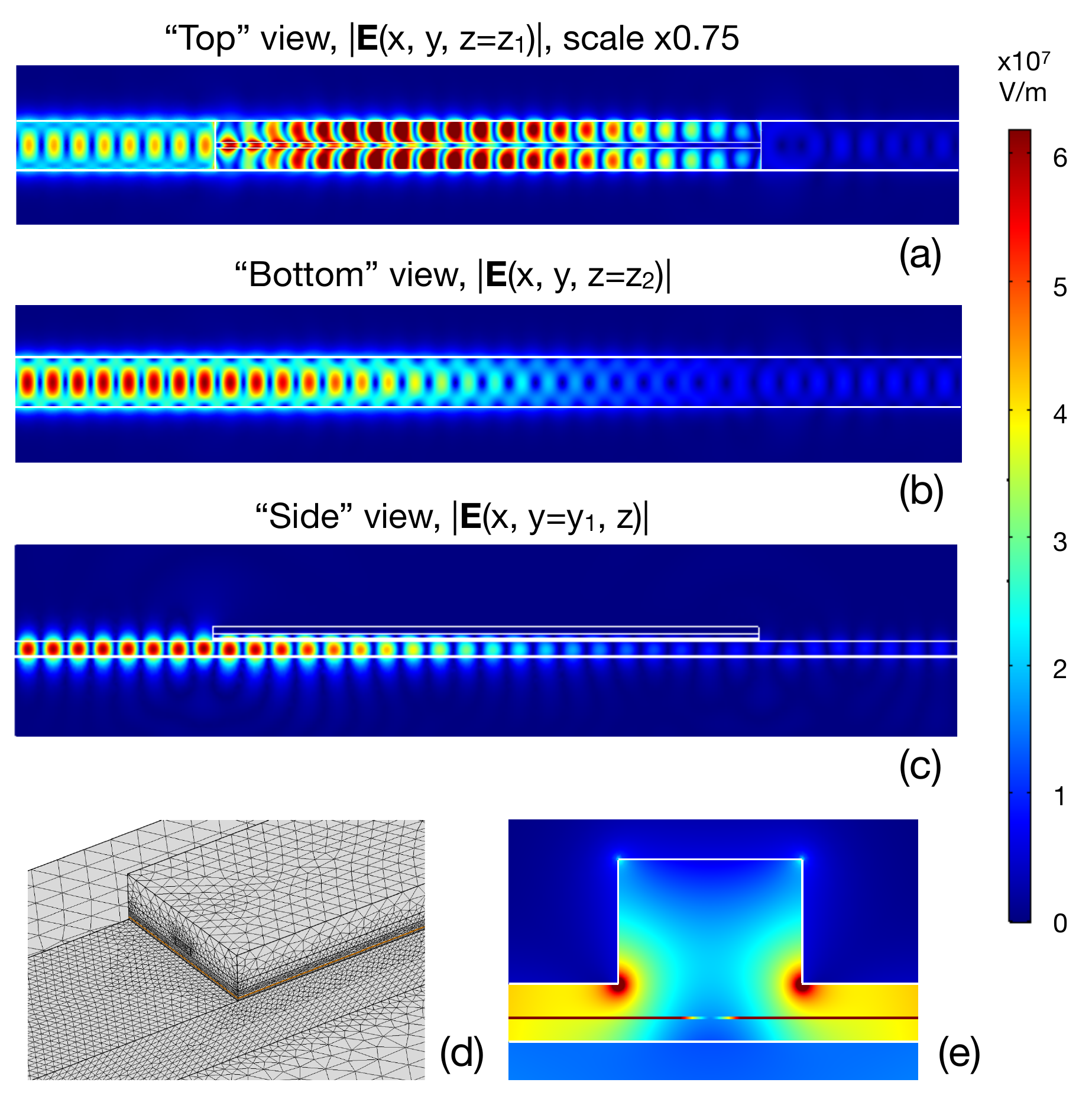}}
\caption {Distribution of the electric field absolute value $|\mathbf{E}(x,y,z)|$ inside the modulator when the voltage $U=-8.5$ V is applied (off-state). As in Fig.~\ref{fig_03} three different cross-sections of the 3D field are shown which correspond to 'top view' (a), 'bottom view' (b) and 'side view' (c). Strongly inhomogeneous mesh which was used in numerical calculations to resolve well the thin accumulation layer is shown in (d). Distribution of the field in a hybrid plasmonic mode formed via the assistance of edge plasmons in a presence of the accumulation layer is shown in (e).  \label{fig_05}}
\end{figure}

The accumulation layer with special ENZ properties is formed at the ITO/HfO$_2$ boundary  as the voltage is applied. The layer does not support its own optical modes due to the small thickness. Despite the fact, that ITO in ENZ regime becomes 'more metallic' in a sense that its charge density becomes larger, it cannot support subwavelength SPP modes at the boundary with a dielectric either. To provide appropriate conditions for plasmon excitation the ITO layer should have a large negative real part of the permittivity, which is not the case here. The interaction of ENZ layer with the existing modes is, thus, dictated by boundary conditions. Since the component of the displacement field vector normal to the surface of ITO should be continuous, the condition $E_2=E_1\varepsilon_1/(\varepsilon_2'+i\varepsilon_2'')$ is fulfilled \cite{qiu-2018}, where $E_1$ and $E_2$ are electric fields outside and inside ENZ layer. If the real part of ITO permittivity $\varepsilon_1'$ passes through zero, a resonance in the local field intensity takes place, with a width defined by the imaginary part of permittivity $\varepsilon_2''$. At the same time $\varepsilon_2''$ defines optical losses in ITO and they grow notably, approximately 65 times, inside the accumulation layer at the ENZ regime. Therefore, an efficient absorption occurs, when the significant part of the field is concentrated inside ITO accumulation layer. An opposite effect appears when we change the polarity of the voltage. In the resulting depletion layer  $\varepsilon_2'$ may be larger than $\varepsilon_1$. In this case ITO pushes the field out of the active layer.

The distribution of the electric field inside the modulator under the applied voltage is shown in Fig.~\ref{fig_05} (a)-(c). The output in this case is significantly weaker than the input. The value of the transmission coefficient is $T_{\text{off}}=0.019$. Using the previously obtained on-state transmission ($T_{\text{on}}=0.747$) we  evaluate the extinction coefficient as
\begin{equation}
\kappa_{\text{ext}}=10\log_{10}{\frac{I_{\text{on}}}{I_{\text{off}}}}=15.95 \, \text{dB}.
\label{extinction_coeff}
\end{equation}
An oscillatory mode behavior, typical for the on-state (Fig.~\ref{fig_03}), cannot be observed for the off-state in Fig.~\ref{fig_05}. The beats do not appear also for the elongated model as it will be shown below. It suggests that the presence of the accumulation layer changes the character of interaction between modes in the modulator and may affect the period of beats.

\begin{figure}
\centerline{\includegraphics[width=0.45\textwidth]{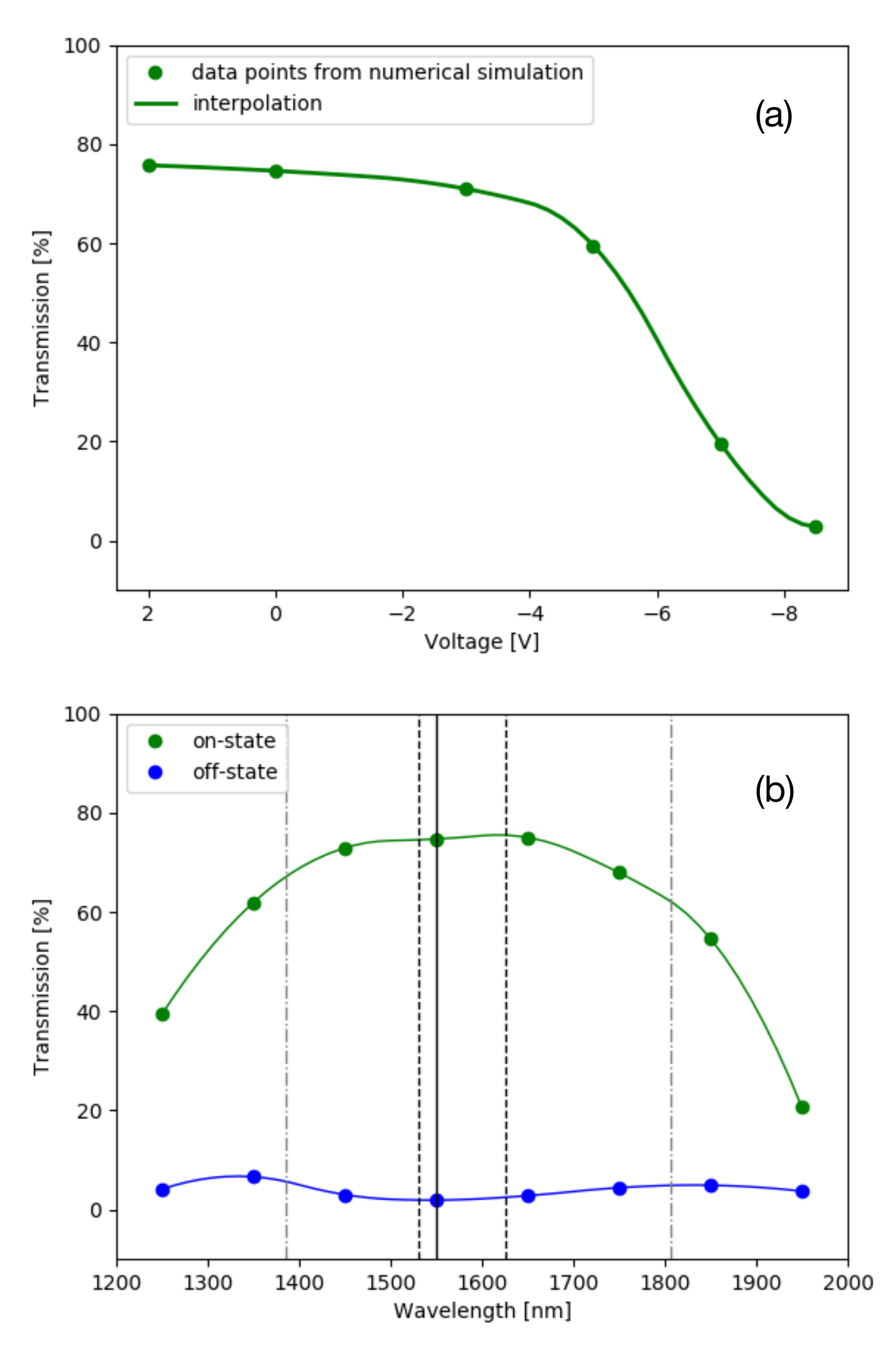}}
\caption {Dependence of the modulator transmission coefficient on the applied voltage (a) and on the wavelength of light for both on-state and off-state (b). Vertical solid black line corresponds to the wavelength $1550$~nm. C+L telecom band is shown using vertical dashed black lines. The optical bandwidth of the modulator, defined in the text, is shown with vertical gray dash-dotted lines. \label{fig_06}}
\end{figure}

\begin{figure*}
\centerline{\includegraphics[width=0.95\textwidth]{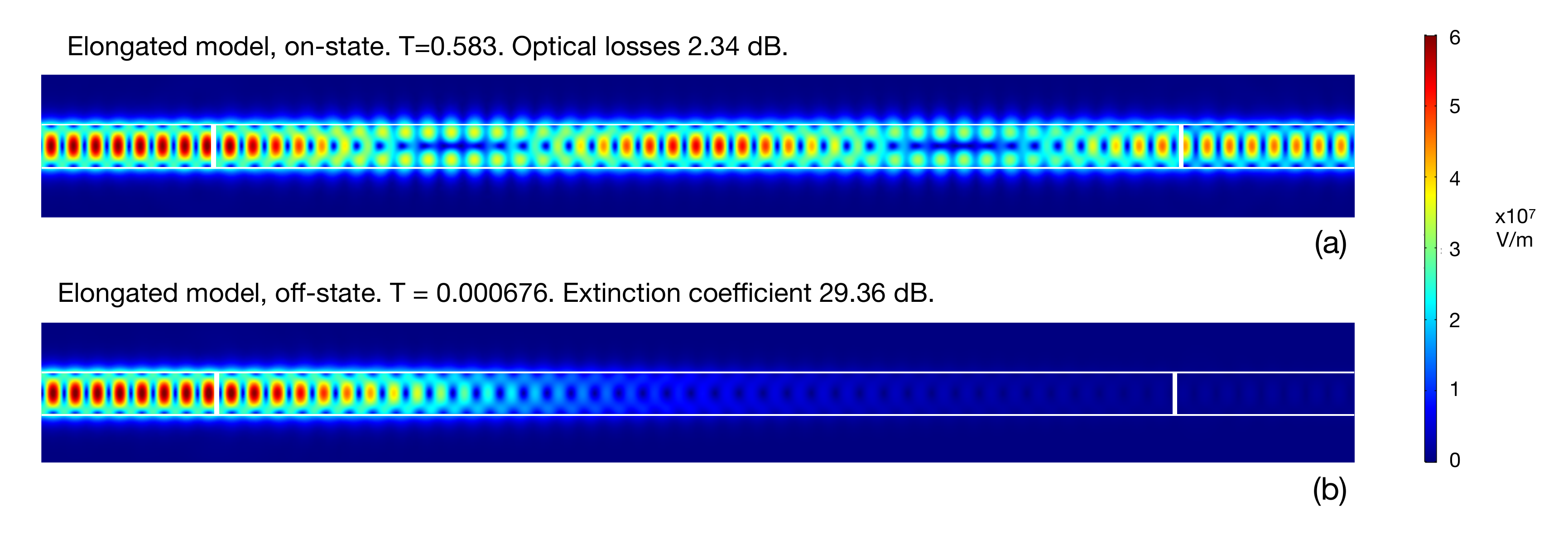}}
\caption {Electric field distribution in the enlarged model (length $L$ of the modulating sandwich is $13.7$ $\mu$m) of the electro-optical modulator. On-state (a) and off-state (b) are demonstrated. Obtained values of optical losses coefficient and extinction coefficient are $2.34$ dB and $29.36$ dB respectively. \label{fig_07}}
\end{figure*}

A small, about $1$ nm, width of the accumulation layer creates a resolution challenge in numerical computations. It is complicated to resolve density and, thus, permittivity variation given in Fig.~\ref{fig_04} (b) in 3D computations, since the profile varies significantly at a small scale. To build an efficient numerical model we used the concept of an effective accumulation layer with constant ENZ parameters placed in a contact with ordinary unperturbed ITO. For a verification of the selected approach we performed 2D mode analysis, where two cases were compared. First, the hybrid plasmonic mode was computed when the electrons density distribution (and corresponding permittivity distribution in ITO) given in Fig.~\ref{fig_04} (b) was used explicitly. Second, we computed the same mode, but using an effective layer with a fixed width and constant ENZ parameters instead of a distribution. Computations were repeated with varying effective layer width until the effective index of the mode and the field distribution became very close to the original computation with continuous permittivity distribution. This procedure can be considered as an alternative way to determine the width of the accumulation layer. It gives $1$ nm thickness, which coincides with the evaluation in Sec.~\ref{sec_transport}. The effective layer concept is used then to build a 3D numerical model. The field distribution in the hybrid mode is shown in Fig.~\ref{fig_05} (e) (note the intensity maximum inside the thin accumulation layer at ITO/HfO$_2$ boundary). Taking into account, that only a small fraction of ITO can be switched to ENZ regime, we obtain the modulator characteristics that are slightly less impressive than those where the whole volume of ITO is considered as active. Nevertheless, our evaluation is realistic and obtained numbers are sufficient for potential applications.

Despite the simplification provided by the effective layer model, it is still a challenge to resolve a $1$ nm thick material in 3D. To compute the field numerically, strongly inhomogeneous mesh is used (see Fig.~\ref{fig_05} (d)). First, the planar triangular mesh is built at the boundary between the insulator and ITO with a variable element size of $10$ - $20$ nm. Then the plane is copied in the directions of HfO$_2$ and ITO ($z$-direction) with steps $3$ nm and $0.3$ nm respectively to reproduce $10$ nm thick insulator and  $1$ nm thick active layer of ITO. After that, a 3D tetrahedral mesh is generated in other areas with variable elements sizes depending on the refractive indices of materials. Resulting mesh contains roughly $6$ million elements with the size varied from $0.3$ nm to $155$ nm. One numerical experiment takes approximately $200$ Gb of RAM and $24$ hours of computational time at HPC cluster of Skoltech.

Additional characteristics of the modulator are presented in Fig.~\ref{fig_06}. The switching process is illustrated in details by a smooth dependence of the transmission on the applied voltage in Fig.~\ref{fig_06}a. Maximal absorption regime (off-state) corresponds to the point where the accumulation layer with ENZ properties is fully formed ($-8.5$ V). Further increase of the voltage does not improve the extinction. Positive voltages allow to form the depletion layer at HfO$_2$/ITO boundary instead of the accumulation layer. Nevertheless, according to Drude theory (see Eq.~\ref{drude}), ENZ effect is not possible in this regime and corresponding variation of the refractive index is much less pronounced. For this reason, strong modulation regime does not exist at positive voltages and we do not consider them. Optical bandwidth of the modulator can be defined using the data in Fig.~\ref{fig_06}b, where the wavelength dependence of the transmission in both on-state and off-state is shown. We define the band, using  Eq.~\ref{extinction_coeff}, as an interval around the central wavelength ($1550$ nm) where the extinction drop is less that $5$ dB. The obtained $421$ nm wide band is shown in Fig.~\ref{fig_06}b using vertical gray dash-dotted lines (at $1385$~nm and $1806$~nm). It is obviously much larger than the C+L telecom band shown using vertical black dashed lines.

The performance and main characteristics of plasmonic electro-optical modulators depend on the length of the modulating sandwich in a nontrivial way. As discussed in Sec.~\ref{sec_design}, the period of a population exchange between the waveguide mode and the plasmonic mode should be consistent with the length of the modulator. Optical computations for the model with $13.7$ $\mu$m long modulating sandwich, i.e. twice the oscillation period, are shown in Fig.~\ref{fig_07}. Both on- and off-state are demonstrated with corresponding optical losses and the extinction coefficients. The transmission coefficient for the doubled model ($T=0.583$) is just $1.3$ times smaller than the corresponding coefficient of $6.8$ $\mu$m long model ($T = 0.747$), which, probably, means that the system loose approximately $25\%$ of the signal per each oscillation period plus $10\%$ for input and output. Optical losses is a typical issue in plasmonic applications and there are different suggestions on how to overcome them in the future \cite{leosson-2012, berini-2012}. Since the attenuation of the signal in the off-state (see Fig.~\ref{fig_07} (b)) is exponential, the enlarged model leads to the significant increase in the modulation depth (the transmission is almost $30$ times smaller than in the $6.8$ $\mu$m long model), which can be an advantage for a certain types of applications. Another advantage of the long model is the decreased active resistance. Since the area of the sandwich $S_{\text{mod}}$ and, consequently, the electric contact area in the $xy$-plane is twice larger, the resistance $R$ is twice smaller, which decreases the $RC_0$ time of the capacitor and makes the modulator bandwidth larger. It is remarkable that $13.7$ $\mu$m long waveguide based electro-optical modulator is still much more compact than many solutions without HPWG. At the same time, low optical losses and predicted THz bandwidth limit of the short model (Fig.~\ref{fig_05}) also look promising for applications.

If the geometry of the sandwich is modified, the character of the interaction between modes changes as well, which influences the period of beats in HPWG. Therefore, the length of the modulator should be optimized for each geometry of the sandwich (taking into account both on- and off-states). Since the numerical optimization is computationally expensive and time consuming, the development of a reasonable analytic theory of the modes interplay in HPWG would be a nice task for the future.

\vspace{-0.8cm}

\section*{Conclusion}  \label{sec_conclusion}

The new approach utilizing edge plasmons in optoelectronics is developed. It allows to couple a horizontally polarized waveguide mode to the plasmonic mode via appropriately designed HPWG. The described idea helps to design compact and efficient electro-optical modulators. The following advantages of the proposed design should be emphasized: (a) the possibility to remove the polarization constraint, thus matching the modulator with recently proposed highly efficient grating couplers \cite{michaels-2018} without the need to use lossy polarization converters; (b) steady planar geometry following from the fact that silicon waveguides for horizontally polarized modes have the width lager than height; (c) consequently, the possibility to keep the active electric resistance $R$ lower; (d) the horizontal polarization makes it possible to put an electrode at the bottom of the waveguide and avoid parasitic plasmon modes excitation. The proposed design implies the usage of a plasmonic rail with two golden ribs at the Au/ITO boundary supporting edge plasmon modes. The design is relatively simple, which is crucial for potential experimental realizations and future applications. To optimize the geometry and structure of the device, numerically heavy 3D optical model based on Maxwell equations is developed. The details of charge density behavior are obtained from the drift-diffusion system of equations. The electron distribution in the accumulation layer at ITO/HfO$_2$ boundary is studied and implemented in the optical computations. The most important characteristics of the device, such as optical losses, the extinction coefficient and the bandwidth are computed. The obtained numbers make the device attractive for potential applications.

\vspace{-0.3cm}

\section*{Acknowledgement}

This work was financially supported by the Ministry of Science and Higher Education of the Russian Federation, project No.RFMEFI58117X0026.

I.A.P. thanks Victor Vysotskiy for the support of Pardus cluster and fruitful discussions.


%


\end{document}